\documentstyle[12pt,cite,epsfig]{article}

\renewcommand{\today}{6 December, 1995}

\newcommand{\nc}{\newcommand}
\nc{\be}{\begin{equation}}
\nc{\ee}{\end{equation}}
\nc{\bea}{\begin{eqnarray}}
\nc{\eea}{\end{eqnarray}}
\nc{\beas}{\begin{eqnarray*}}
\nc{\eeas}{\end{eqnarray*}}
\nc{\noi}{\noindent}
\nc{\sD}{\not \! \! D}
\nc{\s}[1]{\not \! #1}
\nc{\non}{\nonumber}
\nc{\bb}{\bibitem}
\nc{\rw}{$\rho\!-\!\omega$ }
\nc{\rg}{$\rho\!-\!\gamma$ }
\nc{\lf}{\left}
\nc{\r}{\right}
\nc{\mb}[1]{\makebox[#1]{}}
\nc{\pa}{\partial}
\nc{\sA}{\not \! \! A}
\nc{\newsec}[1]{\section{#1}\mb{0.5cm}}
\nc{\h}{\frac{1}{2}}
\nc{\ra}{\rightarrow}
\nc{\la}{\leftarrow}
\nc{\ep}{$e^+e^-\ra\pi^+\pi^-$}
\def\mathunderaccent#1{\let\theaccent#1\mathpalette\putaccentunder}
\def\putaccentunder#1#2{\oalign{$#1#2$\crcr\hidewidth
\vbox to.2ex{\hbox{$#1\theaccent{}$}\vss}\hidewidth}}

\nc{\ti}{\mathunderaccent\tilde}
\nc{\sq}{\sqrt{q^2}}

\begin{document}
\thispagestyle{empty}
\begin{flushright}
ADP-95-49/T196 \\
hep-ph/9510425
\end{flushright}
%\thanks{Thanks to ....}

\begin{center}
%\pagenumbering{arabic}
{\large{\bf  Vector Meson Mixing
and Charge Symmetry Violation}} \\
\vspace{.95cm}
{\large{\bf  [Phys. Lett. B370 (1996) 12-16]}}\\
\vspace{.95cm}
H.B.\ O'Connell$^a$, A.G.\ Williams$^a$, M.\ Bracco$^b$ and G.\ Krein$^b$ \\
\vspace{.25cm}
\vskip24pt
{\it
$^a$Department of Physics and Mathematical Physics, University of Adelaide,\\
S.Aust 5005, Australia}
\vskip24pt
{\it
$^b$Instituto de F\'{\i}sica Te\'orica, Universidade Estadual Paulista,\\
Rua Pamplona 145, 01405-900 S\~ao Paulo--SP, Brazil} \\
\vskip24pt
\today
\vskip24pt
\begin{abstract}

We discuss the consistency of the traditional vector meson dominance (VMD)
model for photons coupling to matter, with the vanishing of vector
meson-meson and meson-photon mixing self-energies at $q^2=0$.  This vanishing
of vector mixing has been demonstrated in the context of rho-omega mixing for
a large class of effective theories. As a further constraint on such models,
we here apply them to a study of photon-meson mixing and VMD.  As an
example we compare the
predicted momentum dependence of one such model with a momentum-dependent
version of VMD discussed by Sakurai in the 1960's.  We find that it produces
a result which is consistent with the traditional VMD phenomenology. We
conclude that comparison with VMD phenomenology can provide a useful
constraint on such models.

\end{abstract}
\end{center}
\vfill
\begin{flushleft}
E-mail: {\it
hoconnel,awilliam@physics.adelaide.edu.au; mirian,gkrein@ift.unesp.br}

\end{flushleft}
\newpage

The experimental extraction of $\Pi_{\rho\omega}$ (in the pion EM form-factor
\cite{Barkov}) is in the timelike $q^2$ region around the \rw mass, yet it is
used to generate charge symmetry violation (CSV) in boson exchange models of
the $NN$ interaction in the spacelike region\cite{HM,MTW}.  The traditional
assumption was that the mixing amplitude was independent of $q^2$.

This assumption was first questioned by Goldman {\it et al.} \cite{GHT} who
constructed a model in which the $\rho$ and $\omega$ mixed via a quark loop
contribution which is non-vanishing if and only if $m_u\neq m_d$.  Their
conclusion of a {significant} momentum dependence was subsequently supported by
other studies, which included an analogous $NN$-loop calculation \cite{PW}
using the $n$-$p$ mass difference and more elaborate quark-loop model
calculations \cite{KTW}. All of these predicted a similar momentum-dependence
for $\Pi_{\rho\omega}(q^2)$ with a node near the origin ($q^2=0$).  At a more
formal level, it was subsequently shown that the vector-vector mixings must
identically vanish at $q^2=0$ in a large class of effective theories
\cite{OPTW} where the mixing occurs exclusively through coupling of the vector
mesons to conserved currents and where the vector currents commute in the usual
way. Recent work in chiral perturbation theory and QCD sum rules has also
suggested that such mixing matrix elements must, in general, be expected to be
momentum dependent \cite{chiral}.

In response to this, alternative mechanisms involving CSV  have been proposed
\cite{Gardner}. Indeed, as the vector mesons are off shell, the individual
mechanisms should not be examined in isolation, because they are dependent on
the choice of interpolating fields for the vector mesons and are not physical
quantities. It has been argued that one could find a set of interpolating
fields for the rho and omega such that {\em all} nuclear CSV occurs through a
constant \rw mixing with the CSV vertex contributions vanishing \cite{cohen}.
However this possibility has been questioned on the grounds of unitarity and
analyticity \cite{kim}.

The same models which have been used to examine the question of \rw mixing
can also be applied to studies of \rg mixing.  They can then be
compared to phenomenology and vector meson dominance (VMD) models, which have
{\em traditionally} assumed the coupling of the photon to the rho was
independent
of $q^2$.  The first person to raise this question was
Miller \cite{Miller}.  The purpose of this letter is to carefully
explore the issues raised and compare numerical predictions for such a mixing
model with experimental data.  As discussed recently \cite{rw_review},
the appropriate representation of VMD to use with a momentum dependent
photon-rho coupling is VMD1, given by the Lagrangian \cite{Sak}
\be
{\cal L}_1=-\frac{1}{4}F_{\mu\nu}F^{\mu\nu}-
\frac{1}{4}\rho_{\mu\nu}\rho^{\mu\nu}
+\h m_\rho^2\rho_\mu\rho^\mu-g_{\rho\pi\pi}\rho_\mu J^\mu-eA_\mu J^\mu
-\frac{e}{2g_\rho}F_{\mu\nu}\rho^{\mu\nu}+\ldots.
\ee
where $J_\mu$ is the hadronic current and $F_{\mu\nu}$ and $\rho_{\mu\nu}$
are the EM and $\rho$ field strength tensors respectively (the dots refer
to the hadronic part of the Lagrangian). {From} this we
obtain the VMD1 expression for the
form-factor for the pion \cite{rw_review}
\be
F^{(1)}_\pi(q^2)=1-
\frac{q^2 g_{\rho\pi\pi}}{g_\rho[q^2-m_\rho^2+im_\rho\Gamma_\rho]}.
\label{FF1}
\ee
Note that in the limit of exact universality $g_{\rho\pi\pi}=g_\rho$ and we
recover the usual VMD2 model prediction for the pion form-factor
\cite{rw_review,Sak}.  Recall that in this traditional VMD (i.e., VMD2)
model the photon couples to hadrons
{\em only} through first coupling to vector mesons with a {\em constant}
coupling strength, e.g., for the $\rho-\gamma$ coupling we have
$\Pi^{\rm VMD2}_{\rho\gamma}(q^2)\equiv -m_\rho^2e/g_\rho$

We shall define a VMD1-like model to be one in which
the photon couples to the hadronic field both
directly and via a $q^2$-dependent coupling (with a node at $q^2=0$) to
vector mesons. A VMD1-like model may differ from pure VMD1 as the coupling of
the photon to the rho (generated by some microscopic process)
will not generally be linear in $q^2$. Hence $g_\rho$, which is a constant
in VMD1 (and VMD2 as they share the same $g_\rho$ \cite{rw_review,Sak}),
may acquire some momentum dependence in a VMD1-like model; the test
for the phenomenological validity of the model is then that this
momentum dependence for $g_\rho$ is not too strong.
For example, we can easily determine the coupling of the
photon to the pion field via the rho meson for a
for a VMD1-like model.  We note the appearance
in Eq.~(\ref{ours}) of the
photon-rho mixing term, $\Pi^{\mu\nu}_{\rho\gamma}(q^2)$, which can be
determined from Feynman rules, and which will, in general, be $q^2$-dependent.
Such an analysis gives for any VMD1-like model
\begin{eqnarray}
-i{\cal M}^{\mu}(q^2)&\equiv&
-ie(p^+-p^-)_{\sigma}[D_\gamma(q^2)]^{\mu\sigma} F_\pi(q^2) \nonumber\\
&=&i[D_{\gamma}(q^2) ]^{\mu\sigma}i[\Gamma_{\gamma\pi}(q^2)]_{\sigma} +
i[D_{\gamma}(q^2)]^{\mu\sigma} i[\Pi_{\gamma\rho}(q^2)]_{\sigma\tau}
i[D_{\rho}(q^2)]^{\tau\nu} i[\Gamma_{\rho\pi}(q^2)]_{\nu} \nonumber\\
&=&-ie(p^+-p^-)_{\sigma} i[D_{\gamma}(q^2)]^{\mu\sigma}\left[1 +
{\Pi_{\rho\gamma}(q^2) \over q^2 -m^2_{\rho}+im_\rho\Gamma_\rho}
{g_{\rho\pi\pi} \over e} \right]\;,
\label{ours}
\end{eqnarray}
where $D$, $\Pi$ and $\Gamma$ denote propagators, one-particle irreducible
mixing amplitudes and proper vertices respectively. Here $p^+$ and $p^-$ are
the outgoing momenta of the $\pi^+$ and $\pi^-$ respectively.  For this model
to reproduce the phenomenologically successful VMD, and hence provide a good
fit to the data (assuming exact universality), $\Pi_{\rho\gamma}(q^2)$ and
$g_\rho$ must be related by (comparing Eqs.~(\ref{FF1}) and (\ref{ours}))
\be
\Pi^{\rm VMD1}_{\rho\gamma}(q^2)
= -\frac{q^2e}{g_\rho(q^2)}.
\label{ident}
\ee
Note that this result then implies that $\Pi^{\rm VMD1}_{\rho\gamma}(q^2) =
(q^2/m_\rho^2) \Pi^{\rm VMD2}_{\rho\gamma}(q^2)$.
Eq.~(\ref{ident}) arises from the simple VMD1 picture when
universality is assumed and is also consistent with the usual VMD2 picture
as explained elsewhere
\cite{rw_review,Sak}.

Thus Eq.~(\ref{ident}) is the central equation of this work, since
vector-meson mixing models (e.g., $\rho-\omega$ mixing) can also be used to
calculate $\rho-\gamma$ mixing and then confronted with traditional VMD
phenomenology.  The results quoted in the review by Bauer et al. \cite{Bauer}
are summarised in Tables I and XXXII of that reference.  They list a range of
values which vary depending on the details of the fit to the $\rho$ mass
($m_\rho$) and width ($\Gamma_\rho$).  Within the context of the traditional
VMD (i.e., VMD2) framework they extract $g_\rho^2(q^2=0)/4\pi$ from $\rho^0$
photoproduction ($\gamma p\to\rho^0 p$) and  $g_\rho^2(q^2=m_\rho^2)/4\pi$ from
$\rho^0\to e^+e^-$.  The three sets of results quoted are (in an obvious
shorthand notation): $\Gamma_\rho =135,145, 155$MeV, $m_\rho=767,774,776$MeV,
$g_\rho^2 (q^2=0)/4 \pi=2.43\pm 0.10,2.27\pm 0.23,2.18\pm 0.22$, $g_\rho^2
(q^2=m_\rho^2)/4 \pi=2.21\pm 0.017,2.20\pm 0.06,2.11\pm 0.06$ respectively.  We
see that $g_\rho$ is a free parameter of the traditional VMD model (VMD2) which
is adjusted to fit the available cross section data.  The central feature of
the VMD2 model is that it presumes a constant value for its coupling constant
$g_\rho$. We note in passing that the universality condition is $g_{\rho}\sim
g_{\rho\pi\pi} \sim g^{\rm univ}_{\rho NN}\sim g_{\rho\rho\rho}$ and where
experimentally we find \cite{Bauer,Dumbrajs} for each of these $g^2/4\pi\sim
2$. For example, the values of $g_{\rho\pi\pi}$ corresponding to the above
three sets of results are $g_{\rho\pi\pi}^2 (q^2=m_\rho^2) /4\pi=2.61,
2.77,2.95$ and are extracted from $\rho^0 \to\pi^+ \pi^-$.  It should be noted
that the $\rho NN$ interaction Lagrangian is here defined as in Refs.\
\cite{MTW,PW} with no factor of two \cite{Sak,Dumbrajs} and hence $g_{\rho
NN}=g^{\rm univ}_{\rho NN}/2$. As a typically used value is $g_{\rho
NN}^2/(4\pi)=0.41$ we see that universality is not accurate
to better than $40\%$ in $g_\rho^2$, which corresponds to $\simeq 20\%$ in
$g_\rho$.

The
results of the VMD2 analysis \cite{Bauer} are approximately consistent with
$g_\rho$ being a constant and so we see from Eq.~(\ref{ident}) that
$\Pi_{\rho\gamma}$ in VMD1-like models should not deviate too strongly from
behaviour linear with $q^2$.

We shall now examine the process within the context of the model used by
Piekarawicz and Williams (PW) who considered \rw mixing as being generated by
a nucleon loop \cite{PW} within the Walecka model.  Using nucleon loops as
the intermediate states removes the formation of unphysical thresholds in the
low $q^2$ region and allows us to use well-known parameters.  The
rho-coupling is not a simple, vector coupling, but rather \cite{Weber}
\be \Gamma^\mu_{\rho
NN}=g_{\rho NN}\gamma^\mu+i\frac{f_{\rho NN}}{2M} \sigma_{\mu\nu}q^\nu,
\ee
where $C_\rho\equiv f_{\rho NN}/g_{\rho NN}=6.1$ and $M$ is the nucleon mass.
With the introduction of tensor coupling the model is no longer
renormalisable, but to one loop order we can introduce some appropriate
renormalisation prescription.  As the mixings are transverse, we write
$\Pi_{\mu\nu}(q^2)=(g_{\mu\nu}-q_\mu q_\nu/q^2)\Pi(q^2)$ \cite{OPTW}.  The
photon couples to charge, like a vector and so, unlike the PW calculation, we
have only a proton loop to consider.
Here we can safely neglect the coupling of the photon to the nucleon magnetic
moment and so there is no neutron loop contribution nor any tensor-tensor
contribution to the proton loop.
This sets up two kinds of mixing, vector-vector
$\Pi^{\mu\nu}_{\rm vv}$ and vector-tensor $\Pi^{\mu\nu}_{\rm vt}$, where
(using dimensional regularisation with the associated scale, $\mu$)
\bea
\Pi_{\rm vv}(q^2)&=&-q^2\frac{eg_{\rho NN}}{2\pi^2}
\left[\frac{1}{3\epsilon}-\frac{\gamma}{6}-\int_0^1dx\;x(1-x)\ln\left(
\frac{M^2-x(1-x)q^2}{\mu^2}\right)\right], \label{pi1} \\
\Pi_{\rm vt}(q^2)&=&-q^2\frac{eg_{\rho NN}}{8\pi^2}
\left[\frac{1}{\epsilon}-\gamma-\int_0^1dx\;\ln\left(
\frac{M^2-x(1-x)q^2}{\mu^2}\right)\right].
\label{pi2}
\eea
Note that these functions vanish at $q^2=0$, as expected from the node
theorem since we have coupling to conserved currents \cite{OPTW}.  To remove
the divergence and scale-dependence we add a counter-term \[{\cal L}_{CT}
=e\frac{g_{\rho NN} C_T}{2\pi^2} \rho_{\mu\nu}F^{\mu\nu}\] to the Lagrangian
in a minimal way so as to renormalise
the model to one loop.  This will contribute
$-iC_Tg_{\rho NN}eq^2/\pi^2$ to the photon-rho vertex, which will add to the
contribution $i\Pi$ generated by the nucleon loop. The counter-term will
contain pieces proportional to $1/\epsilon$, $\gamma$ and $\ln \mu^2$ to
cancel the similar terms in Eqs.~(\ref{pi1}) and (\ref{pi2}), and a constant
piece, $\beta$, which will be chosen to fit the extracted value for
$g_\rho(0)$. The counter-term is
\be
C_T=-\frac{1}{\epsilon}\left(\frac{1}{6}+\frac{C_\rho}{8}\right)+\gamma\left(
\frac{1}{12}+\frac{C_\rho}{8}\right)-\left(\frac{1}{12}
+\frac{C_\rho}{8}\right)\ln\mu^2+\beta,
\ee
which gives
us the renormalised mixing,
\bea
\non
\Pi_{\rho\gamma}(q^2)&=&q^2\frac{eg_{\rho NN}}{\pi^2}\left[\frac{1}{2} \left(
\frac{5}{18}+\frac{2M^2}{3q^2}-\frac{8M^4+2M^2q^2-q^4}{3q^3\sqrt{4M^2-q^2}}
\arctan\sqrt{\frac{q^2}{4M^2-q^2}}\right.\right.\\
&-&\left.\frac{\ln M^2}{6}\right)
\left.+\frac{C_\rho}{8}\left(-2+2\sqrt{\frac{4M^2-q^2}{q^2}}\arctan
\sqrt{\frac{q^2}{4M^2-q^2}}+\ln M^2\right) -\beta\right].
\eea
We find that the choice $\beta=8.32$ in our counter-term
approximately reproduces the extracted
value of $g_\rho(0)$ at $q^2=0$.

\begin{figure}[htb]
  \centering{\
     \epsfig{angle=0,figure=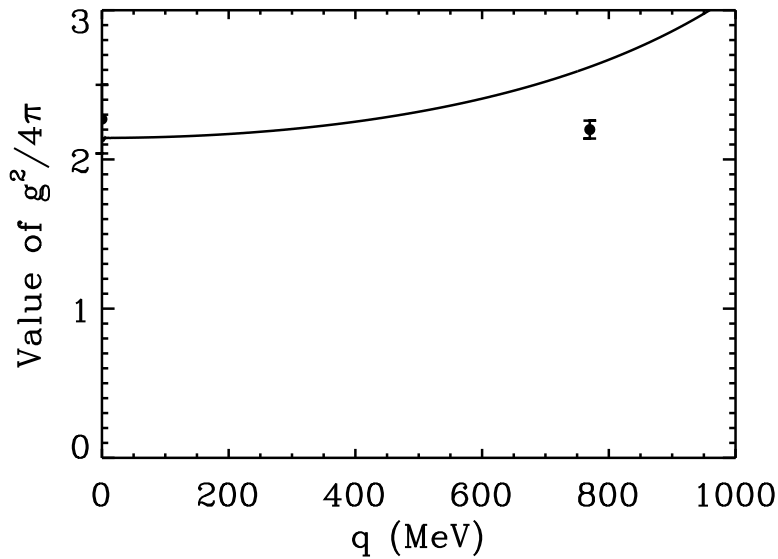,height=8.5cm}
               }
\parbox{130mm}{\caption
{The PW model prediction for the mixing amplitude is related to
the traditional VMD coupling $g_\rho(q^2)$
using the central result of Eq.~(\protect{\ref{ident}}).  The resulting
behaviour of $g_\rho^2/4\pi$ versus $q\equiv \protect{\sq}$ is then plotted
in the timelike region for this model.
Shown for comparison are a typical pair of results
($2.27\pm 0.23$ at $q=0$ and $2.20\pm 0.06$ at $q=m_\rho$, see text) taken
from a traditional VMD based analysis of cross section
data in Ref.\ \protect{\cite{Bauer}}.
}
\label{graph}}
\end{figure}

The results for $g_\rho(q^2)$ for the PW model are shown in Fig.~\ref{graph}.
We see that despite this model having a node in the photon-rho mixing at
$q^2=0$ the resulting $q^2$ dependence of $g_\rho$ is small.
As can be seen from this plot, we obtain values of $g_\rho^2(0)/(4\pi)=
2.14$ and $g_\rho^2(m_\rho^2)/(4\pi)=2.6$ compared to the
experimental averages 2.3 and 2.17 respectively.

It should be remembered that Eq.~(\ref{ident}) is only as reliable as
universality, which is itself violated at a level of 30-40\% . Based
on this important observation, we can conclude then that the PW model provides
a result consistent with the spread of extracted results given in
Ref.~\cite{Bauer}. It should be noted that any VMD1-like model which predicts a
significantly greater deviation from linearity with $q^2$ will fail to
reproduce phenomenology because of Eq.~(\ref{ident}).

In summary, we have explicitly shown in Eq.~(\ref{ident}) that the vanishing of
vector-vector mixing at $q^2=0$ is completely consistent with the standard
phenomenology of vector meson dominance (VMD).  We have, in addition, applied
the same type of model used in a study of \rw mixing to extract the momentum
dependence of \rg mixing and have compared the result to the VMD2 based
analysis of the experimental data.  We see that the phenomenological
constraints of VMD can provide a useful independent test of VMD-like models of
vector mixing and future studies should take adequate account of this.  It
would, of course, be preferable to reanalyse the data used in Ref.~\cite{Bauer}
from the outset using VMD1 rather than VMD2, but this more difficult task is
left for future investigation.

\vspace{1.5cm} {\bf Acknowledgments}: AGW and HOC would like to thank A.W.\
Thomas for helpful discussions and a careful reading of the manuscript.  AGW
acknowledges the hospitality of the IFT, S\~{a}o-Paulo, Brazil where part of
this work was carried out.  This work was supported in part by FAPESP and
CNPq (Brazil) and the Australian Research Council.

%\newpage

\end{document}